\pdfoutput=1
\RequirePackage{ifpdf}
\ifpdf 
\documentclass[pdftex]{sigma}
\else
\documentclass{sigma}
\fi

\numberwithin{equation}{section}

\begin{document}

\allowdisplaybreaks

\renewcommand{\PaperNumber}{015}

\FirstPageHeading

\ShortArticleName{On a~Trivial Family of Noncommutative Integrable Systems}

\ArticleName{On a~Trivial Family of Noncommutative
\\
Integrable Systems}

\Author{Andrey V.~TSIGANOV}

\AuthorNameForHeading{A.V.~Tsiganov}

\Address{St.~Petersburg State University, St.~Petersburg, Russia}
\Email{\href{mailto:andrey.tsiganov@gmail.com}{andrey.tsiganov@gmail.com}}

\ArticleDates{Received October 17, 2012, in f\/inal form February 18, 2013; Published online
February 22, 2013}

\Abstract{We discuss trivial deformations of the canonical Poisson brackets associated with the
Toda lattices, relativistic Toda lattices, Henon--Heiles, rational Calogero--Moser
and~Ruijsenaars--Schneider systems and
apply one of these deformations to construct a~new trivial family of noncommutative integrable
systems.}

\Keywords{bi-Hamiltonian geometry; noncommutative integrable systems}

\Classification{37J35; 53D17; 70H06}

\section{Introduction}

Let us consider some smooth manifold $M$ with coordinates $x_1,\ldots,x_m$ and~a~dynamical system
def\/ined by the following equations of motion
\begin{gather*}
\dot{x}_i=X_i,\qquad i=1,\ldots,m.
\end{gather*}
We can identify this system of ODE's with the vector f\/ield
\begin{gather*}
X=\sum X_i\dfrac{\partial}{\partial x_i},
\end{gather*}
which is a~linear operator on a~space of the smooth functions on $M$ that encodes the
inf\/initesimal evolution of any quantity
\begin{gather*}
\dot{F}=X(F)=\sum X_i\dfrac{\partial F}{\partial x_i}.
\end{gather*}
In Hamiltonian mechanics one of the fundamental axiom is what we can call the energy paradigm that
can be stated as follows:
``For every mechanical system there is a~function def\/ined on its space of states, called
mechanical energy or Hamiltonian $H$ of the system, containing all its dynamical information''.

According to this paradigm any function $H$ on $M$ generates vector f\/ield $X$ describing
a~dynamical system
\begin{gather*}
X=X_H=PdH.
\end{gather*}
Here $dH$ is a~dif\/ferential of $H$, and~$P$ is a~bivector on the phase space $M$.
By adding some other assumptions we can prove that $P$ is a~Poisson bivector.
In fact, it is enough to add energy conservation
\begin{gather*}
\dot{H}=X_H(H)=(PdH,dH)=0
\end{gather*}
and compatibility of dynamical evolutions associated with two functions $H_{1,2}$
\begin{gather*}
X_{H_1}(X_{H_2}(F))=X_{H_2}(X_{H_1}(F))+X_{X_{H_1}(H_2)}(F),
\end{gather*}
see~\cite{ib96,jo64} and references therein.

In bi-Hamiltonian mechanics~\cite{mag97} we are looking for another decomposition of the given
vector f\/ield $X$,
\begin{gather*}
X=PdH=f_1P'dH_1+\cdots+f_mP'dH_m,
\end{gather*}
by commuting Hamiltonian vector f\/ields
\begin{gather*}
X_k=P'dH_k
\end{gather*}
generated by integrals of motion $H_1,\ldots,H_m$ and~some Poisson bivector $P'$ compatible with
$P$.
These Poisson bivectors $P'$ can be divided into two groups of trivial and~nontrivial deformations
of canonical Poisson bivector $P$, see~\cite{al11,mag05,lih77,mag97,ts12a} and~references therein.

Supposing that $P'$ is a~trivial deformation of canonical Poisson bivector $P$ completely def\/ined
by Hamiltonian $H$ we can join the geometry of the phase space $M$ and~the energy paradigm.
In this case, second Poisson bivector
\begin{gather}
\label{poi-2}
P'=\mathcal L_{Y} P
\end{gather}
is a~Lie derivative of the canonical bivector $P$ along the Liouville vector f\/ield
\begin{gather}
\label{y1}
Y=\operatorname{Ad} H,
\end{gather}
where $A$ is a~2-tensor f\/ield acting on the dif\/ferential of the Hamiltonian.
Remind, that the Lie derivative $P'$~\eqref{poi-2} is a~trivial deformation because it is
2-coboundary and~simultaneously 2-cocycle in the Poisson--Lichnerowicz cohomology def\/ined by
canonical Poisson bivector $P$~\cite{mag05,lih77}.

The main aim of this note is to show 2-tensor f\/ields $A$ associated with some well-known
integrable systems and~to prove that these tensor f\/ields may be useful to construct new
integrable systems.
For example, we discuss a~new trivial family of integrable noncommutative three dimensional
systems, which includes deformations of the rational Calogero--Moser system with three particle
interaction.

\section{Integrable systems on cotangent bundles}
\label{section2}

Let us consider canonical Poisson bivector on the symplectic manifold $M=T^*Q$
\begin{gather}
\label{can-p}
P=\sum_{i=1}^n \frac{\partial}{\partial q_i}\wedge \frac{\partial}{\partial p_i},
\end{gather}
which is the one mostly used in Hamiltonian mechanics~\cite{am78,arn89}.
Here $q_i$ are local coordinates describing a~point $q$ on a~smooth manifold $Q$, and~$p_i$, the
canonical conjugate momenta, are local coordinates describing covectors on such manifold,
i.e.\;points $p$ in the cotangent bundle $T^*Q$ of $Q$.
The corresponding Poisson bracket looks like
\begin{gather}
\label{can-br}
\{p_i,q_j\}=\delta_{ij},\qquad \{q_i,q_j\}=\{p_i,p_j\}=0.
\end{gather}

In local coordinates $x$ on $M$ the Lie derivative of a~bivector $P$
along a~vector f\/ield $Y$ reads as
\begin{gather*}
\bigl(\mathcal L_{Y}P\bigr)_{ij}=\sum\limits_{k=
1}^{\dim M}\left(Y_k\dfrac{\partial P_{ij}}{\partial x_k}
-P_{kj}\dfrac{\partial Y_i}{\partial x_k}-P_{ik}\dfrac{\partial Y_j}{\partial x_k}\right)
\end{gather*}
and the Schouten bracket $[A,B]$ of two bivectors $A$ and~$B$
is a~trivector with the following entries
\begin{gather}
\label{sh-br}
[A,B]_{ijk}=-\sum\limits_{m=1}^{\dim M}\left(B_{mk}\dfrac{\partial A_{ij}}{\partial x_m}
+A_{mk}\dfrac{\partial B_{ij}}{\partial x_m}+\mathrm{cycle}(i,j,k)\right).
\end{gather}
In our case $M=T^*Q$, $\dim M=2n$ and~$x=(q,p)$.

Let us consider natural Hamilton functions on $M=T^*Q$
\begin{gather}
\label{nat-h}
H=T(q,p)+V(q),
\end{gather}
which are the sum of the geodesic Hamiltonian $T$ and~potential energy $V(q)$.
According to~\cite{gts11,mpt11,ts10,ts11s}, for natural Hamiltonians there is other representation
for the vector f\/ield $Y$~\eqref{y1}
\begin{gather*}
Y=\left(
\begin{matrix}
\Lambda&0
\\
0&\Pi
\end{matrix}
\right)\left(
\begin{matrix}
dq
\\
dp
\end{matrix}
\right).
\end{gather*}
Here we use matrix notation of tensor objects in which, for instance, canonical Poisson bi\-vec\-tor~$P$~\eqref{can-p} looks like
\begin{gather}
\label{can-p2}
P=\left(
\begin{matrix}
0 & E
\\
-E & 0
\end{matrix}
\right).
\end{gather}
Here $E$ is a~unit matrix.

If $\Pi=0$ and~$\Lambda$ is the conformal Killing tensor of gradient type or Yano--Killing tensor on~$Q$, one gets second Poisson bivector $P'=\mathcal L_YP$
associated with Hamilton functions separable in orthogonal coordinate systems on~$Q$.
In this case eigenvalues of the
the Nijenhuis operator
\begin{gather}
\label{rec-op}
N=P'P^{-1},
\end{gather}
which is also called the hereditary or recursion operator, are variables of separation.
In order to get integrals of motion $H_k$ from $N$
we have to extend the initial phase space~\cite{imm00} or to f\/ix separated relations~\cite{ts10}.

The new idea is that we can substitute the arbitrary Hamilton function $H$~\eqref{nat-h}
and the 2-tensor f\/ield $A$ into the def\/inition~\eqref{poi-2} and
try to f\/ind the Poisson bivectors $P'$ solving the equation
\begin{gather}
\label{sch-eq}
[P',P']=[\mathcal L_{\operatorname{Ad} H}P,\mathcal L_{\operatorname{Ad} H}P]=0,
\end{gather}
where $[\cdot,\cdot]$ is the Schouten bracket def\/ined by~\eqref{sh-br}.
In this case $P'$ will be the Poisson bivector compatible with $P$ and~we will say that $M=T^*Q$ is
the bi-Hamiltonian manifold~\cite{mag97}.
The next step is a~search of integrals of motion for this Hamilton function $H$.

If the recursion operator
$N$ at every point has $n$ distinct functionally
independent eigenvalues, we can say that $M$ is a~regular bi-Hamiltonian manifold.
If the recursion operator $N$ does
not have this property then we can say that bi-Hamiltonian manifold $M$ is
{irregular}~\cite{mpt11, ts10}.
So, there are three dif\/ferent cases:
\begin{enumerate}\itemsep=0pt
\item[1)] recursion operator produces the necessary number of integrals of motion;
\item[2)] recursion operator generates variables of separation instead of integrals of motion;
\item[3)] recursion operator produces only part of the integrals of motion or variables of
separation.
\end{enumerate}
In the third case we have to complement the recursion operator with some additional information in
order to get integrals of motion.
Namely this property allows us to get noncommutative
integrable systems, which will be considered in Section~\ref{section3}.

Now let us show a~collection of tensor f\/ields $A$ associated with some well-known integrable
systems.

\subsection{Toda lattice}
\label{section2.1}

Let us consider the following tensor f\/ield $A$ depending only on $q$ variables
\begin{gather*}
A=\left(
\begin{matrix}
B+2D-B^\top&0
\\
0&B-B^\top
\end{matrix}
\right)
P,
\end{gather*}
where $B$ is a~strictly upper diagonal matrix
\begin{gather}
\label{b-mat}
B=\left(
\begin{matrix}
0 & 1 & 1 & \cdots & 1
\\
0 & 0 & 1 & \cdots & 1
\\
\vdots & & & \ddots & \vdots
\\
& & &0 & 1
\\
0 & & \ldots & & 0
\end{matrix}
\right)=
\sum_{i>j}^n e_{ij},
\end{gather}
and $D$ is a~diagonal matrix
\begin{gather*}
D=\operatorname{diag}(q_1,q_2,\ldots,q_n)=\sum_{i=1}^n q_i e_{ii}.
\end{gather*}
Matrices $ e_{ij}$ are $n\times n$ with only one non zero $(ij)$ entry, which equals to unit.

Substituting this tensor f\/ield $A$ and~a~natural Hamilton function
\begin{gather*}
H=\sum_{i=1}^n p_i^2+V(q)
\end{gather*}
into the def\/inition of $P'$ one gets a~system of equations~\eqref{sch-eq} on $V(q)$.
One of the partial solutions of this system is the Hamilton function for the open Toda lattice
associated with $\mathcal A_n$ root system
\begin{gather*}
H=\sum_{i=1}^n p_i^2+a\sum_{i=1}^{n-1}e^{q_i-q_{i+1}},\qquad a\in\mathbb R.
\end{gather*}
Traces of powers of the corresponding recursion operator $N$~\eqref{rec-op}
\begin{gather}
\label{tr-ham}
H_k=\operatorname{tr} N^k, \qquad k=1,\ldots,n,
\end{gather}
are functionally independent constants of motion in bi-involution
with respect to both Poisson brackets
\begin{gather*}
\{H_i,H_j\}=\{H_i,H_j\}'=0.
\end{gather*}
This Poisson bivector $P'$ was found by Das, Okubo and~Fernandes~\cite{das89,fern93}.

In generic case we can use a~more complicated tensor f\/ield
\begin{gather*}
\widetilde{A}=\left(
\begin{matrix}
\widetilde{B}+\widetilde{D}&0
\\
0&\widetilde{C}
\\
\end{matrix}
\right)P,
\end{gather*}
where entries of $\widetilde{D}$ are linear on $q_i$ and~$\widetilde{B}$ and~$\widetilde{C}$ are
numerical matrices.
Here $\widetilde{B}$ is an arbitrary matrix, whereas $\widetilde{A}$ and~$\widetilde{B}$ satisfy to
algebraic equations which may be obtained from~\eqref{sch-eq} at $V(q)=0$.

Using this tensor f\/ield $\widetilde{A}$ we can get the recursion operators which produce either
integrals of motion for the periodic Toda lattice~\cite{gts06} or variables of separation for the
Toda lattice~\cite{ts07}.
In similar manner we can consider the Toda lattices associated with other classical root
systems~\cite{ts10}.

\subsection{Relativistic Toda lattice}
\label{section2.2}

If we substitute the Hamilton function $H(q,p)$ and~the following tensor f\/ield $A$
\begin{gather*}
A=\left(
\begin{matrix}
-B^\top&0
\\
-E&0
\end{matrix}
\right)
P=\left(\begin{matrix}
0&-B^\top
\\
0&-E
\end{matrix}
\right),
\end{gather*}
where $E$ is a~unit matrix and~$B$ is given by~\eqref{b-mat},
into the def\/inition of $P'$~\eqref{poi-2} we will obtain a~system of equations on $H$.
One of the solutions is the Hamiltonian of the open discrete Toda lattice associated with $\mathcal
A_n$ root system
\begin{gather}
\label{rtl-ham}
H=\sum_{i=1}^n \bigl( c_i+d_i \bigr),
\end{gather}
where $c_i$ and~$d_i$ are the so-called Suris variables
\begin{gather*}
c_i=\exp(p_i-q_i+q_{i+1}),\qquad d_i=\exp(p_i),\qquad
q_0=-\infty,\qquad q_{n+1}=+\infty.
\end{gather*}
Traces of powers of the corresponding recursion operator $N$~\eqref{tr-ham} are integrals of
motion in bi-involution with respect to both Poisson brackets.
Namely this Poisson bivector~$P'$~\eqref{poi-2} is discussed in~\cite{rag89,sur93}.

Remind, that according to~\cite{sur93} there is an equivalence between the relativistic Toda lattice
and the discrete time Toda lattice.
Namely, substituting
\begin{gather*}
p_j =\theta_j +\frac12 \ln\left(
{\frac{1+\exp({q}_j-{q}_{j-1} )}{1+\exp({q}_{j+1} -{q}_j )}}\right)
\end{gather*}
in~\eqref{rtl-ham} one gets standard Hamiltonian for the
relativistic Toda lattice
\begin{gather*}
H=\sum_{j=1}^{n-1} \exp (\theta_j )
\Bigl[\bigl[1+\exp({q}_j -{q}_{j-1} ) ]
[ 1+\exp({q}_{j+1} -{q}_j ) \bigr]\Bigr]^{1/2}.
\end{gather*}
Transformation $(\theta_j\cdot q_j )\rightarrow
(p_j\cdot q_j )$ is a~canonical transformation.

As above, two numerical matrices $\widetilde{B}$ and~$\widetilde{C}$ in the tensor f\/ield
\begin{gather*}
A=\left(
\begin{matrix}
0&\widetilde{B}
\\
0&\widetilde{C}
\\
\end{matrix}
\right)
\end{gather*}
allow us to get recursion operators $N=P'P^{-1}$ which generate either integrals of motion
for the periodic relativistic Toda lattice or variables of separation~\cite{kuz94}.

\subsection[Henon-Heiles system]{Henon--Heiles system}
\label{section2.3}

At $n=2$ we can introduce the following linear in momenta tensor f\/ield $A$
\begin{gather}
\label{a-hh}
A=\left(
\begin{matrix}
B & 0
\\
0 & C
\\
\end{matrix}
\right)P=\left(
\begin{matrix}
0 & B
\\
-C & 0
\\
\end{matrix}
\right),
\end{gather}
where
\begin{gather*}
B=\left(
\begin{matrix}
2q_1p_1&q_1p_2
\\
q_1p_2&q_2p_2
\\
\end{matrix}
\right),\qquad C=\left(
\begin{matrix}
f_1(q)p_1+f_2(q)p_2&0
\\
0&f_3(q)p_1+f_4(q)p_2
\\
\end{matrix}
\right).
\end{gather*}
Substituting this tensor f\/ield $A$ and~a~natural Hamilton function
\begin{gather*}
H_1=p_1^2+p_2^2+V(q)
\end{gather*}
into the def\/inition of $P'$ one gets a~system of equations~\eqref{sch-eq} on $V(q)$ and~functions
$f_k(q)$.
The resulting system of PDE's has two partial polynomial solutions
\begin{gather*}
V(q)={c_1} q_2\big(3q_1^2+16q_2^2\big)+c_2 \left(2q_2^2
+\dfrac{q_1^2}{8}\right)+c_3q_2,\qquad c_k\in\mathbb R,
\end{gather*}
and
\begin{gather*}
V(q)={c_1}\big(q_1^4+6q_1^2q_2^2+8q_2^4\big)
+{c_2}\big(q_1^2+4q_2^2\big)+\dfrac{c_3}{q_2^2}.
\end{gather*}
Second integrals of motion $H_2 =\operatorname{tr} N^2$ are fourth order polynomials in momenta.

So, one gets the Henon--Heiles potential and~the fourth order potential~\cite{gdr84} as
particular polynomial solutions of the equations~\eqref{sch-eq} associated with tensor
f\/ield~\eqref{a-hh}.

Using slightly deformed tensor f\/ield $A$ we can get the same systems with singular
terms~\cite{gts11} and
their three-dimensional counterparts~\cite{ts10}.

\subsection[Rational Calogero-Moser model]{Rational Calogero--Moser model}
\label{section2.4}

Following~\cite{mpt11} let us consider tensor f\/ield $A$, which is proportional to $P$
\begin{gather*}
A=\rho(q,p)P,
\end{gather*}
where $\rho(q,p)$ is a~function on $M$.
If
\begin{gather}
\label{a-cal}
A=(p_1q_1+\cdots+p_nq_n)P, \qquad \rho=p_1q_1+\cdots+p_nq_n,
\end{gather}
then equations~\eqref{sch-eq} have the following partial solution
\begin{gather}
\label{ham-cal}
H=\dfrac{1}{2}\sum_{i=1}^{n}p_{i}^{\;2}+\dfrac{g^2}{2}\sum_{i\neq
j}^n\frac{1}{(q_{i}-q_{j})^{2}},
\end{gather}
where $g$ is a~coupling constant.
It is the Hamilton function of the $n$-particle rational Calogero--Moser model associated with the
root system $\mathcal A_n$.

The corresponding recursion operator $N$~\eqref{rec-op} generates only a~Hamilton function
\begin{gather*}
\operatorname{tr}N^k=2H^{k},\qquad k=1,\ldots,n,
\end{gather*}
that allows us to identify our phase space $M=\mathbb R^{2n}$ with the irregular bi-Hamiltonian
manifold.

In this case~\cite{mpt11,ts10} integrals of motion are polynomial solutions of the equations
\begin{gather}
\label{cal-int}
P dH=-\frac{1}{k} P' d\ln H_k,\qquad k=1,\ldots,n,
\end{gather}
which have two functionally independent solutions for any $k\geq2$.
It is easy to see that the functions
\begin{gather}
\label{caz-cal}
C_{km}=\frac{H_m^{-1/m}}{H_k^{-1/k}}
\end{gather}
are Casimir functions of $P'$, i.e.~$P'dC_{km}=0$.

Some solutions of equations~\eqref{cal-int} coincide with the well-known integrals of motion
\begin{gather*}
J_{n-m}\equiv\frac{1}{m!}\underbrace{\bigg\{\sum_{i=1}^n q_{i}\cdots\bigg\{
}_{m~\text{times}}\sum_{i=1}^n q_{i},J_{m}\bigg\}\cdots\bigg\},\qquad m=1,\dots,n-1,
\end{gather*}
obtained from the conserved quantity
\begin{gather}
J_{n}\equiv \exp\left(-\frac{g^{2}}{2}\sum_{i\neq j}
\frac{1}{(q_{i}-q_{j})^{2}}\frac{\partial^{2}}{\partial
p_{i}\partial p_{j}}\right)\prod_{k=1}^{n}p_{k}
\nonumber
\end{gather}
by taking its successive Poisson brackets with $\sum\limits_{i=1}^{n}q^{i}$~\cite{gon01}.
These $n$ solutions, including $J_2=H$, are in involution with respect to the Poisson
brackets~\eqref{can-br}.

Other $n-1$ functionally independent solutions of~\eqref{cal-int},
\begin{gather*}
K_{m}=m g_{1}J_{m}-g_{m}J_{1},\qquad
g_{m}=\frac{1}{2}\left\{\sum_{i=1}^{n}q_{j}^{\;2},J_{m}\right\},\qquad
m=2,\dots,n,
\end{gather*}
are not in involution with respect to the canonical Poisson bracket def\/ined
by~\eqref{can-p}~\cite{gon01}.

\subsection[Rational Ruijsenaars-Schneider model]{Rational Ruijsenaars--Schneider model}
\label{section2.5}

Let us consider tensor f\/ield $A$, which is proportional to canonical bivector $P$
\begin{gather*}
A=(q_1+\cdots+q_n)P,\qquad\rho=q_1+\cdots+q_n.
\end{gather*}
In this case equations~\eqref{sch-eq} have the following partial solutions
\begin{gather}
\label{tr-rtl}
J_k=\frac{1}{k!}\operatorname{tr} L^k,\qquad k=\pm 1,\pm2,\ldots,\pm n,\end{gather}
where $L$ is the Lax matrix of the Ruijsenaars--Schneider model
\begin{gather*}
L=\sum_{i,j=1}^n\frac{\gamma}{q_i-q_j+\gamma} b_j e_{ij},\qquad
b_k=e^{p_k}\prod_{j\neq k}\left(1-\frac{\gamma^2}{(q_k-q_j)^2}\right)^{1/2}.
\end{gather*}
As above recursion operator produces only the Hamilton function.
It is easy to prove that traces of powers of the Lax matrix $L$
\eqref{tr-rtl} satisfy to the following relations
\begin{gather}
\label{rs-rel}
P dJ_{\pm 2}=-\frac{1}{k} P' d\ln J_k,\qquad k=\pm 1,\ldots,\pm n,
\end{gather}
instead of the standard Lenard--Magri relations~\cite{mag97,tt12}.
Moreover, similar to the Calogero--Moser system, there are other solutions $K_m$ of these
equations~\eqref{rs-rel}, which are described in~\cite{afg12}.

Remind, that the so-called principal Ruijsenaars--Schneider Hamiltonian has the form
\begin{gather*}
H_{RS}=\frac{1}{2}(J_1+J_{-1})=\sum_{k=
1}^n(\cosh2p_k)\prod_{j\neq k}\left(1-\frac{\gamma^2}{(q_k-q_j)^2}\right)^{1/2}
\end{gather*}
and that the rational Ruijsenaars--Schneider system is in duality with the corresponding variant of
the trigonometric Sutherland system, see~\cite{afg12} and  references therein.

We want to highlight that for all integrable systems listed in~\cite{mpt11,tt12,ts10,ts11s}
the second Poisson bivector $P'$~\eqref{poi-2} is a~Lie derivative of the canonical Poisson
bivector $P$ along the vector f\/ield
$Y=\operatorname{Ad}H$~\eqref{y1}, where tensor f\/ield $A$ usually has a~very simple form.

In the next section we show that such simple tensor f\/ields $A$ may be useful to search for new
integrable systems.

\section{Noncommutative integrable systems}
\label{section3}

The extreme rarity of integrable dynamical systems makes the quest for them all the more exciting.
We want to apply tensor f\/ields $A$ to partial solution of this problem.
Below we present a~method to construct a~new family of three dimensional noncommutative integrable
systems.

Let us consider natural Hamilton function on $M=\mathbb R^{2n}$
\begin{gather*}
H=\sum_{i=1}^n p_i^2+V(q_1,\ldots,q_n)
\end{gather*}
and bivector $A$ associated with the rational Calogero--Moser system~\eqref{a-cal}
\begin{gather*}
A=(p_1q_1+\cdots+p_nq_n)P,
\end{gather*}
where $P$ is canonical Poisson bivector~\eqref{can-p}, \eqref{can-p2}.

In previous section we have discussed partial solutions of the equations~\eqref{sch-eq}, here we
want to discuss their complete solution.
\begin{proposition}
The Lie derivative of $P$~\eqref{can-p} along the vector field $Y$
\begin{gather}
\label{p-p}
P'=\mathcal L_Y P,\qquad Y=(p_1q_1+\cdots+p_nq_n)PdH
\end{gather}
is a~Poisson bivector compatible with $P$ if and~only if
\begin{gather}
\label{m-pot}
H=\sum_{i=1}^n p_i^2+\frac{1}{q_1^2}
F\left(\frac{q_2}{q_1},\frac{q_3}{q_1},\ldots,\frac{q_n}{q_1}\right).
\end{gather}
Here $F$ is an arbitrary homogeneous function of zero degree function depending on the homogeneous
coordinates
\begin{gather*}
x_1=\frac{q_2}{q_1},\quad x_2=\frac{q_3}{q_1},\quad\ldots,\quad x_{n-1}=\frac{q_n}{q_1}.
\end{gather*}
\end{proposition}

The definition of the homogeneous coordinates may be found in~\cite{gr94}.
Proof is a~straightforward calculation of the Schouten bracket~\eqref{sch-eq}.

It is easy to see that some Hamilton functions separable in spherical coordinates and~Hamilton
functions for the rational Calogero--Moser systems associated with the $A_n$, $B_n$, $C_n$
and~$D_n$ root systems have the form~\eqref{m-pot}.

We got accustomed to believing that the notion of two compatible Poisson structures $P$ and~$P'$
allows us to get the appropriate integrable systems~\cite{gts11,mag97,tt12,ts10,ts11s}.
In our case recursion operator $N=P'P^{-1}$ reproduces only the Hamilton function
\begin{gather*}
\operatorname{tr}N^k=2(2H)^{k}.
\end{gather*}
It allows us to identify our phase space $M=\mathbb R^{2n}$ with the irregular bi-Ha\-mil\-to\-nian
mani\-fold~\mbox{\cite{mag97,ts10}}, but simultaneously it makes the use of standard constructions of the
integrals of motion impossible.

We do not claim that all the Hamilton functions~\eqref{m-pot} are integrable because we do not have
an explicit construction of the necessary number of integrals of motion.
Nevertheless, even in generic case there is one additional integral of motion.
\begin{proposition}
The following second order polynomial in momenta
\begin{gather*}
C=(p_1q_1+\cdots+p_nq_n)^2-(q_1^2+\cdots+q_n^2)H
\end{gather*}
is a~Casimir function of $P'$, i.e.~$P'dC=0$.
\end{proposition}
Consequently we have
\begin{gather*}
\{H,C\}=0.
\end{gather*}
It is enough for integrability at $n=2$ when we get Hamilton functions
\begin{gather*}
H=p_1^2+p_2^2+\frac{1}{q_1^2} F\left(\dfrac{q_2}{q_1}\right)
\end{gather*}
separable in polar coordinates on the plane.

At $n>3$ we can make some assumptions on the form of the additional integrals of motion.
For instance,
let us postulate that our dynamical system is invariant with respect to translations, i.e.\;that
there is a~linear in momenta integral of motion
\begin{gather*}
H_{\rm post}=p_1+\cdots+p_n,\qquad\{H,H_{p\rm ost}\}=0.
\end{gather*}
It leads to the additional restriction on the form of the proper Hamilton functions~\eqref{m-pot}
\begin{gather*}
H=\sum_{i=1}^n p_i^2+\frac{1}{(q_2-q_1)^2} G\left(\frac{q_3-q_2}{q_2-q_1},\frac{q_4-q_3}{q_2-q_1},\ldots,\frac{q_n-q_{n-1}}{q_1-q_2}\right),
\end{gather*}
which generate bi-Hamiltonian vector f\/ields
\begin{gather}
\label{vf-biham}
X=PdH=P'd\ln H_{\rm post}^{-1}
\end{gather}
equipped with the four integrals of motion
\begin{gather}
\label{4-int}
H_1=H_{\rm post},\qquad H_2=H,\qquad H_3=C,\qquad H_4=\{H_1,C\}
\end{gather}
with the linearly independent dif\/ferentials $dH_i$.
According to the Euler--Jacobi theorem~\cite{jac36} it is enough for integrability by quadratures
at $n=3$.

Remind that the Euler--Jacobi theorem~\cite{jac36} states that a~system of $N$ dif\/ferential
equations
\begin{gather}
\label{eqm}
\dot{x}_i=X_i(x_1,\ldots,x_N),\qquad i=1,\ldots,N,\end{gather}
possessing the last Jacobi multiplier $\mu$ (invariant measure) and~$N-2$ independent f\/irst
integrals is integrable by quadratures.
In our case $N=6$, we have four independent integrals of motion~\eqref{4-int} and~$\mu=1$.

{\sloppy So, at $n=3$ the following Hamilton functions
\begin{gather}
\label{3g-ham}
H_2=p_1^2+p_2^2+p_3^3+\frac{1}{(q_2-q_1)^2} G\left(\frac{q_3-q_2}{q_2-q_1}\right)
\end{gather}
labelled by functions $G$ generate integrable by quadratures Hamiltonian equations of
motion~\eqref{vf-biham}--\eqref{eqm}.
Because
\begin{alignat}{4}
& \{H_1,H_2\}=0,\qquad&& \{H_1,H_3\}=H_4,\qquad &&\{H_1,H_4\}=2H_1^2-6H_2, &
\nonumber\\
& \{H_2,H_3\}=0,\qquad && \{H_2,H_4\}=0,\qquad &&\{H_4,H_3\}=4H_1H_3 &\label{alg-int}
\end{alignat}
we have noncommutative integrable systems with respect to the canonical
Poisson bracket, see, for instance,~\cite{kh10} and~references therein.

}

Of course, in the center of momentum frame, the total linear momentum of the system is zero $H_1=0$
and~we have three integrals of motion $H_2$, $H_3$ and~$H_4$ in the involution that is enough for
integrability at $n=3$ and~$n=4$.

On the other hand, Hamilton functions~\eqref{3g-ham} def\/ine superintegrable systems in the
Liouville sense
\begin{gather*}
\{H_i,H_j\}'=0,\qquad i,j=1,\ldots,4,
\end{gather*}
with respect to the second Poisson bracket $\{\cdot,\cdot\}'$ associated with the Poisson tensor
$P'$~\eqref{p-p}.
If we put
\begin{gather*}
G(x)=g^2\left(1+\frac{1}{x^2}+\frac{1}{(1+x)^2}\right),
\end{gather*}
we can obtain a~well-known Hamiltonian for the rational Calogero--Moser system~\eqref{ham-cal}
\begin{gather*}
H_{\rm CM}=\sum_{i=
1}^{3}p_{i}^{2}+\frac{g^2}{(q_2-q_1)^2}+\frac{g^2}{(q_3-q_2)^2}+\frac{g^2}{(q_3-q_1)^2}.
\end{gather*}
In this case there are other polynomial integrals of motion~\eqref{cal-int} and~other Casimir
functions of~$P'$~\eqref{caz-cal}.
This system was separated by Calogero~\cite{cal69} in cylindrical coordinates in $\mathbb R^3$.
Being a~superintegrable system it is actually separable in four other types of coordinate
systems~\cite{bcr00}.
These variables of separation may be easily found using either the generalised Bertrand--Darboux
theo\-rem~\cite{rw03} or methods of the bi-Hamiltonian geometry~\cite{gts05}.
Remind, that variables of separation are eigenvalues of the Killing tensor $K$ satisfying equation
\begin{gather}\label{killing-eq}
KdV=0,
\end{gather}
where $V$ is potential part of the Hamiltonian $H$.

Any additive deformation of this function $G(x)$ leads to the integrable additive deformation of
the rational Calogero--Moser system, for instance, if
\begin{gather*}
\widetilde{G}(x)=G(x)+\frac{a}{x},
\end{gather*}
then one gets an integrable system with the three-particle interaction
\begin{gather*}
\widetilde{H}_{\rm CM}=H_{\rm CM}+\frac{a}{(q_1-q_2)(q_2-q_3)}.
\end{gather*}
For this Hamilton function we couldn't f\/ind any polynomial in momenta
integrals of motion except $H_1$, $H_3$ and~$H_4$~\eqref{4-int}).
Moreover, we couldn't get variables of separation using the standard (regular) methods such as
generalised Bertrand--Darboux theorem~\cite{rw03} and~bi-Hamiltonian algorithm discussed
in~\cite{gts05}.
Namely, in contrast with the case $a=0$ at $a\neq0$ the Killing ten\-sor~$K$
satisfying~\eqref{killing-eq} has only functionally dependent eigenvalues.

At $n=3$ in order to get rational Calogero--Moser systems associated with other classical root
systems and~their deformations we can postulate an existence of the fourth order integral of motion
\begin{gather*}
H_{\rm post}=\sum_{i\neq j}^n p_i^2p_j^2+\sum_{k}f_k(q)p_k^2+g(q),
\end{gather*}
with some unknown functions $f_k(q)$ and~$g(q)$.
However we do not have
an exhaustive classif\/ication as of yet.

In generic case at $n\geq3$ we can use other hypotheses about
additional integrals of motion commuting with $H$~\eqref{m-pot}.

Of course, construction of such integrable systems is trivial and~closely related with construction
of the
group invariant solutions of partial dif\/ferential equations through imposing side
conditions~\cite{or87,ov82}.
Remind, we can look for solution $W(q)$ of the Hamilton--Jacobi equation
\begin{gather*}
H(p,q)=\sum_{i=1}^n\left(\frac{\partial W}{\partial q_i}\right)^2+W(q_1,\ldots,q_n)=
\mathcal E,\qquad p_j=\dfrac{\partial W}{\partial q_j},
\end{gather*}
up to the side condition
\begin{gather*}
\mathcal S(q,p)=0.
\end{gather*}
If $\{H,\mathcal S\}=f(q,p)\mathcal S$ then this side condition is consistent with $H$ and~the
corresponding integrals of motion are def\/ined modulo $\mathcal S=0$, i.e.
\begin{gather*}
\{H,H_k\}=g_k(q,p)\mathcal S.
\end{gather*}
Here $f$ and~$g_k$ are some functions on phase space and~$W(q)$ is the so-called characteristic
Hamilton function.

In our case the side condition is related with the transition to the center of momentum frame
\begin{gather*}
\mathcal S=H_{\rm post}\equiv p_1+p_2+p_3=0,
\end{gather*}
which is always consistent with the Hamilton function~\eqref{3g-ham} and~we have three integrals of
motion~$H_2$,~$H_3$ and~$H_4$~\eqref{4-int} in involution by modulo $\mathcal S=0$~\eqref{alg-int}.
Construction of the variables of separation for the Hamilton--Jacobi equation with a~side
conditions is discussed in~\cite{mil12}.

In quantum case we can consider the Schr\"{o}dinger equation
\begin{gather*}
H\Psi=\mathcal E\Psi,\qquad H=\Delta+V(q_1,\ldots,q_n),
\end{gather*}
where $\Delta$ is the Laplace--Beltrami operator on $M=\mathbb R^{2n}$, and~study solution of this
equation that also satisf\/ies a~side condition
\begin{gather*}
\mathcal S\Psi=0.
\end{gather*}
The consistency condition for the existence of nontrivial solutions $\Psi$ is a~standard
\begin{gather*}
[H,\mathcal S]=f\mathcal S.
\end{gather*}
In this case linear dif\/ferential operator $K$ will be a~symmetry operator for $H$ modulo
$\mathcal S\Psi=0$ if
\begin{gather*}
[H,K]=g\mathcal S.
\end{gather*}
Here $f$ and~$g$ are some linear partial dif\/ferential operators, see examples and~discussion
in~\cite{mil12}.

We assume that the quantum counterpart of $H$~\eqref{3g-ham} could
be embedded in this generic scheme.

\section{Conclusion}
\label{section4}

We have demonstrated that the trivial deformations of the canonical Poisson bracket
associated with the well-known integrable systems have a~very simple form def\/ined
by some 2-tensor f\/ield $A$ acting on the dif\/ferential of the Hamilton function.
We have shown
a collection of examples and~also proven that such tensor f\/ields may be useful for searching new
integrable by quadratures dynamical systems.
For example, we have proven the
noncommutative integrability of a~new generalisation of the
rational Calogero--Moser system with three particle interaction.

In fact, we propose a~new form for the old content and~believe that this unif\/ication
is a~next step in creating the invariant and~rigorous geometric theory of integrable systems on
regular and
irregular bi-Hamiltonian manifolds.

\subsection*{Acknowledgements}
We would like to thank E.G.~Kalnins, W.~Miller, Jr.\
and G.~Rastelli for useful discussion on noncommutative integrable systems.
The study was supported by the ministry of education and~science of Russian Federation, project
07.09.2012 no.~8501, grant no.~2012-1.5-12-000-1003-016.

\pdfbookmark[1]{References}{ref}
\LastPageEnding

\end{document}